\documentstyle[11pt,moriond,epsfig]{article}

\def\Journal#1#2#3#4{{#1} {\bf #2}, #3 (#4)}

\def\PLB{{\em Phys. Lett.}  B}
\def\PRL{\em Phys. Rev. Lett.}
\def\PRD{{\em Phys. Rev.} D}

\def\ra{\rightarrow}

\def\be{\begin{equation}}
\def\ee{\end{equation}}
\def\bea{\begin{eqnarray}}
\def\eea{\end{eqnarray}}

\begin{document}
\vspace*{4cm}
\title{HEAVY FLAVOURS}

\author{ S. BARSUK }

\address{Laboratoire de l'Acc\'el\'erateur Lin\'eaire,  
Universit\'e Paris-Sud 11, B\^atiment 200, 91898 Orsay, France}

\maketitle
\begin{figure}[h]
\begin{center}
%\framebox[35mm]{\rule[-11mm]{0mm}{35mm}}
\epsfig{figure=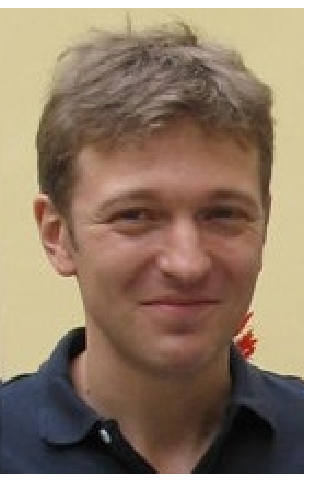,height=35mm}
\end{center}
\end{figure}
\abstracts{
The paper introduces the selection 
of new results on heavy flavours presented 
at the QCD and High Energy Interactions section 
of the XLIIIth ``Rencontres de Moriond'' conference. 
}

New results on heavy flavours~\footnote{
Despite many new important results on $\tau$ and particularly 
top quark physics~\cite{top}, 
we concentrate below on selected topics of the charm and beauty 
sectors.} 
come predominantly from experiments at $e^+ e^-$ machines, as B-factories or CLEOc, 
and from the Tevatron $p \bar{p}$ collisions, exploited by the CDF and D0 
experiments. 
The asymmetric B-factories take data in the region of $\Upsilon$ 
resonances at $\sqrt s \sim 10 \, GeV / c^2$ with prolific production of $B$ mesons 
and reconstruction of their production and decay vertices. 
The integrated luminosity at B-factories adds up to 1.3 $ab^{-1}$ 
corresponding to the statistics of about $10^{9}$ $B_{u,d}$ mesons. 
Despite unfortunate abandoning of PEPII and the BaBar experiment operation, 
the joint accumulated luminosity by the end of 2009 is projected to reach 2 $ab^{-1}$. 
The CLEOc experiment was taking data at $\sqrt s$ of 3 to 5 $GeV / c^2$ 
until spring 2008, having collected unprecedented sample 
of charmonium region data, e.g. few $10^7$ $\psi ( 2S )$ decays. 
The Tevatron experiments have acquired a 3 $fb^{-1}$ integrated 
luminosity each at $\sqrt s$ of 2 $TeV / c^2$, and are expecting to collect 
around 8 $fb^{-1}$ each before 2010. 

New experiments are entering the data taking phase this year, 
BESIII at $e^+ e^-$ machine BEPC with $\sqrt s$ of 3 to 5 $GeV / c^2$, 
and the experiments at the LHC, $pp$ machine 
with $\sqrt s$ of 14 $TeV / c^2$, where the LHCb experiment is 
most promising for the new precision data on charm and beauty physics. 
An annual yield of $10^{12}$ $b \bar{b}$ pairs ($2 fb^{-1}$ of data) 
is expected at LHCb with all the $b$ species produced.
% having the biggest samples of $B_u$, $B_d$, $B_s$, $B_c$ and $\Lambda_b$, 
% and also other baryons and numerous excited states. 
More projects will come later, SuperBelle, SuperB factory 
and SuperLHC(b) will constitue the superfuture of heavy flavour physics. 

The experiments are aimed at precise determination of the Standard Model (SM) parameters, 
and particularly search for any indication 
of the effects beyond the very successful SM picture. 
New physics (NP) can manifest itself via new particles or new couplings and 
is searched either via direct production by increasing the center of mass 
energy as at LHC (ATLAS, CMS) or ILC, or in loops by increasing the 
luminosity as at the B-factories, LHC(b) and SuperB projects. 

\begin{center}
\begin{figure}[h]
\begin{minipage}[b]{7.5cm}
\centering
\epsfig{figure=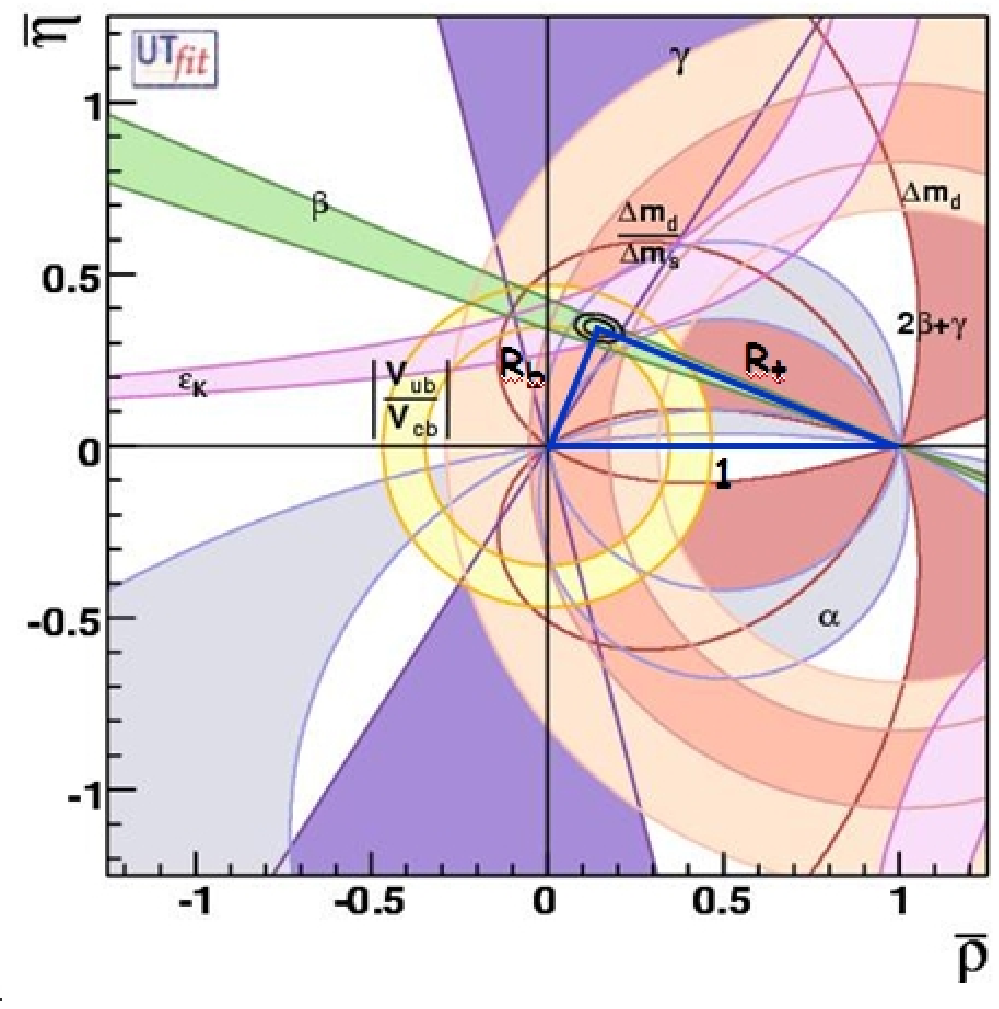,height=5.5cm,width=7.cm}
\caption{The unitarity triangle constraints$^2$}
%\caption{The unitarity triangle constraints~\cite{utfit}}
% from the UTFit group}
\label{fig:fig1}
\end{minipage}
\hspace{0.5cm}
\begin{minipage}[b]{7.5cm}
\centering
\epsfig{figure=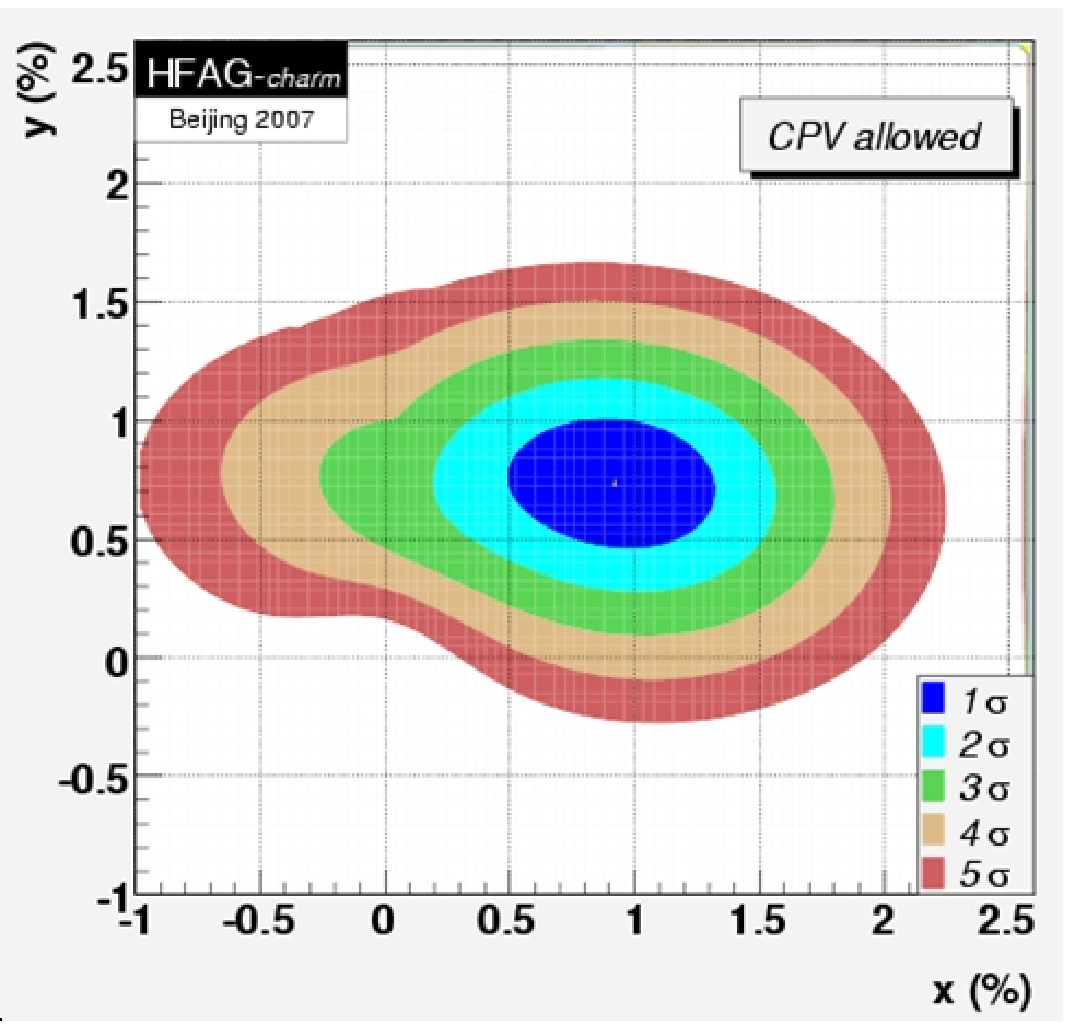,height=5.5cm,width=7.cm}
\caption{Summary of the $D^o \bar{D}^o$ mixing data$^7$}
%\caption{Summary of the $D^o \bar{D}^o$ mixing data~\cite{hfag}}
% by the HFAG group}
\label{fig:fig2}
\end{minipage}
\end{figure}
\end{center}

Essential tests are provided by the studies of non-squashed 
unitarity triangles (UT), visualising the Cabibbo-Kobayashi-Maskawa (CKM) matrix 
unitarity, with the area (Jarlskog invariant) quantifying the $CP$ violation 
as described by SM. 
Owing to the results from B-factories and Tevatron, 
precision of the UT parameters has significantly 
improved, the UT apex is precisely constrained~\cite{utfit} (Fig.~\ref{fig:fig1}). 
The angles are known to the precision of 
$\Delta \alpha \approx 8^o$, 
$\Delta \beta \approx 1^o$ and 
$\Delta \gamma \approx 13^o$, 
where $\beta$ and $\gamma$ are still dominated by experimental error. 
The precision on the sides is dominated by theoretical uncertainties. 
The $R_b$ side determination suffers from theoretical uncertainty 
of $\sim 8\%$ on the $V_{ub}$ extraction, 
while the $R_t$ side ($| V_{td} / V_{ts} |$) 
is known to a precision of $\sim 5\%$. 
In both cases limitations come from lattice calculations, 
some $R_t$ improvement is expected from radiative penguin decay studies. 
The precision of lattice calculations in $B$ sector is improving, 
and the errors go down by a factor $\sim$2 in present 
calculations~\cite{davies}. 
Essential test of the lattice calculations reliability comes from 
the charm sector~\cite{kronfeld}. 
A presently 3$\sigma$ difference between the $D_s$ meson form factor $f_{D_s}$ 
calculations~\cite{follana} and the recent precise CLEOc result~\cite{stone}, 
if confirmed could signify either calculation problem or a hint of a NP 
contribution. 
Given the biggest error still comes from the CLEOc statistics, the updated result 
with the available doubled statistics and improved analysis technique is awaited. 
Comparing the precision of UT angle determination to the precision 
of the opposite side, we notice, that constraining the apex with the 
$( \beta , R_b )$ is limited by the $R_b$ precision, while constraining 
the apex with $( \gamma , R_t )$ is limited by the precision on $\gamma$. 
Present knowledge of the $R_t$ requires the angle 
$\gamma$ to be measured to a precision of $5^o$. 
\vspace*{-0.5cm}

Complementing the direct searches for NP, the UT language 
illustrates the receipt to search for NP in loops, 
by e.g. comparing tree-mediated processes, which 
are thought to be free of NP effects, to those involving loops. 
Comparing the $2 \beta^{eff}$ angle determined via the 
tree-mediated processes, e.g. $B_d \ra J / \psi K_S$, to that from 
the na\"{i}ve average of the decays involving $b \ra s$ transitions, 
where NP can enter the penguin loop,  
e.g. $B_d \ra \phi K_S$, HFAG~\cite{hfag} quotes a 2.2$\sigma$ tension. 
The same comparison in the $B_s$ sector will be done by LHCb 
using tree-mediated $B_s \ra J / \psi \phi$ and 
pure penguin $B_s \ra \phi \phi$ processes. 
The sensitivity to the corresponding difference 
for the $B_s \bar{B}_s$ mixing phase is expected~\cite{leslie} to be 
$\delta_{2 \chi} \sim 6^o$ in one year of nominal operation. 
The summary of hints for NP is discussed in~\cite{zwicky}, 
and the NP $CP$ parity violation (CPV) is discussed in~\cite{hou}. 

The UTfit group~\cite{utfit} pointed out~\cite{ut} that including recent 
results~\cite{cdf,dnol} 
from the CDF and D0 experiments on the time-dependent tagged angular analysis 
of $B_s \ra J / \psi \phi$ decays, in the combined analysis of all the available 
experimental information, the fit preferred value of $-20^o$ 
for the $B_s \bar{B}_s$ mixing phase 
deviates by 3$\sigma$ from the SM value of $-2^o$. 
This outlines the importance of achieving experimetally the sensitivity 
of the SM expectation value, which will be done by LHCb 
with~\cite{lphis} $\sigma_{stat} ( \phi_s ) \sim 1^o$ achieved in one nominal 
year of data taking. 
% 2 $fb^{-1}$. 

Systems of neutral mesons provide another source to search for NP, 
having all possible combinations for 
$x = \Delta m / \Gamma$ and $y = \Delta \Gamma / 2 \Gamma$ of mixing parameters, 
$x,y \sim O(1)$ for $K \bar{K}$, 
$x,y \ll 1$ for $D \bar{D}$, 
$x \sim 1, y \ll 1$ for $B_d \bar{B}_d$, 
$x >> 1, y \sim O(0.1)$ for $B_s \bar{B}_s$ system. 
Experimentally most difficult cases are fast $B_s \bar{B}_s$ 
oscillations that are difficult to resolve, 
and slow $D \bar{D}$ oscillations that are difficult to detect. 
The discovery of the $B_s \bar{B}_s$ mixing~\cite{bsmixing} at the Tevatron 
has proven that $B$ physics can be successfully studied at hadron machines. 
However the ratio of hadronic parameters calculated with lattice 
remains the limiting factor on the $R_t$ side extraction precision.
Few methods have been employed to search for the $D \bar{D}$ mixing. 
The wrong sign $D^o$ decays can be found by comparing the initial flavour 
known e.g. from the charge of pion from $D^{*+} \ra D^o \pi^+$, 
to the flavour determined from the $D$ decay products. 
Measurement of the time dependence gives a separation between the doubly 
Cabibbo suppressed decay and the decay via mixing. 
The analysis of e.g. $D^o \ra K^+ \pi^-$ events provides a measurement of 
simultaneously the $x^{\prime 2}$ and $y^\prime$ parameters, where 
$x^\prime$ and $y^\prime$ are the mixing parameters rotated 
over the strong interaction phase. 
Semileptonic decays $D^o \ra X l^- \bar{\nu}$ provide a special clean case 
where no time analysis is needed, 
and measure $(x^{\prime 2} + y^{\prime 2}) / 2 = (x^2 + y^2) / 2$.
Measuring the difference between the lifetimes of $D^o$ mesons decaying 
via $D^o \ra K^- K^+$ and $D^o \ra K^- \pi^+$ modes provides 
a difference between the $\Gamma_+$ and $\Gamma_-$ of the $CP$ eigenstates, 
and thus measures $y$. 
Finally the time dependent Dalitz analysis using 
$D^o \ra K_S \pi \pi$ yields $x^{\prime 2}$ and $y^\prime$. 
Combining all the available results, 
the $D \bar{D}$ mixing is considered to be established~\cite{hfag} 
% now by combining the results from Belle, BaBar and CLEO experiments, 
(Fig.~\ref{fig:fig2}).  
However the $x$ and $y$ parameters are still to be determined. 
Having established the $D \bar{D}$ mixing, it becomes possible to search 
for CPV effects in the interference between decay and mixing, 
present sensitivity~\cite{charmcp} being of the order of 1\%. 
SM predicts CPV in charm sector to be small, $O( 10^{-3})$. 
%$\sim \frac{2 \eta A^2 \lambda^5}{\lambda} \sim O( 10^{-3})$. 
A $O(1 \%)$ signal of CPV would already signify the NP contribution. 
Simple signature of CPV could be extracted from the chain 
$\psi(3770) \ra D^o \bar{D}^o \ra CP( \pm ) CP( \pm )$ at e.g. CLEOc and BES.  

A 24 $fb^{-1}$ luminosity accumulated by Belle at the $\Upsilon (5S)$ resonance 
gives access to $B_s$ physics with a sample of about 
$3 \times 10^6$ $B_s$ mesons. 
Thus radiative penguin decay $B_s \ra \phi \gamma$ was observed for 
the first time~\cite{bsphigam} with $18 \pm 6$ signal events and 
$BR = ( 5.7 \pm^{1.8}_{1.5} \pm^{1.2}_{1.7} ) \times 10^{-5}$
in agreement with the SM expectation. 
A $10^4$ reconstructed $B_s \ra \phi \gamma$ events in one year of $2 fb^{-1}$ 
is expected~\cite{lbphigam} at LHCb, 
%with $S/B_{bb} > 0.4 @ 90 \% CL$
which will also study other rare $B_s$ decays. 
Owing to clean collected data and established analysis technique 
Belle also obtained~\cite{bsphigam} a limit on rare 
$B_s \ra \gamma \gamma$ decay, sensitive to many NP approaches, 
$BR( B_s \ra \gamma \gamma ) < 8.6 \times 10^{-6} @ 90 \% CL$, 
close to the SM value of $10^{-6}$. 

(Semi-)leptonic $B$ decays involving $\tau$ lepton are 
expected to be sensitive to NP contributions. 
Clean experimental conditions at the $e^+ e^-$ machines, 
fully understood detectors and well-established analysis techniques 
allowed B-factories to explore rare decays with large ``missing energy''. 
First the method was applied to observe~\cite{taunu,taunub} 
$B^- \ra \tau^- \bar{\nu}_\tau$ decay,  
thus measuring the $B$ meson decay constant $f_B$. 
Furthermore given the improved precision on both experimental value and 
lattice value~\cite{ltaunu} for the $f_B$ is achieved, 
the measurement will provide a sensitivity to NP (e.g. $H^{\pm}$) 
contribution to the annihilation diagram of $B^- \ra \tau^- \bar{\nu}_\tau$ 
transition. 
The technique then has been extended to 
$B \ra D^{(*)} \tau \nu_\tau $ decays~\cite{dtaunu,dtaunub}, 
being also promising to search for $B \ra K^{(*)} \nu \bar{\nu}$ 
using higher, e.g. SuperB, statistics. 
Similarly interesting are the rare transitions $B \ra X_s l^+ l^-$~\cite{huber}, 
a $10^4$ reconstructed $B \ra K^* \mu^+ \mu^-$ decays are expected~\cite{lkstmumu} 
in one year at LHCb. 

Over the last years large number of charmonium-like states have been 
discovered using both a conventional 
energy scan~\cite{cleoscan1,cleoscan2}, 
and the initial state radiation (ISR) method successfully employed by 
B-factories~\cite{isr}. 
Thus the whole $\sqrt s$ interval is accessible due the continuous ISR spectrum, 
and the $\alpha_{em}$ suppression is compensated by the important statistics 
of B-factories. 
The interpretation of the states is not clear, 
among popular approaches is to consider the $X(3872)$~\cite{x3872} 
and $Z(4430)$~\cite{z4430} states as four-quark or 
molecular candidates~\cite{mol}. 
The $Y(4260)$~\cite{y4260} 
and its partner $Y(4320)$~\cite{y4320} are hybrid candidates~\cite{hyb}. 
In each case there are extra states nearby. 
The effects of thresholds and mixing between states can 
complicate interpretation~\cite{thresh}. 
Despite remarkable theoretical efforts, there is no 
unique model to explain the states observed. 
Additional information can be gathered by exploring the $\Upsilon$ region 
to search for the corresponding bottomonium-like states. 

%So far the measurements are entirely consistent with the SM predictions. 

\section*{Acknowledgments}
It is my pleasure to thank Ivan Belyaev, Andrey Golutvin, 
Marie-H\'{e}l\`{e}ne Schune, Viola Sordini and Achille Stocchi 
for their help in preparing the talk.

\end{document}